\documentclass[12pt,preprint]{aastex}
\usepackage{epsfig}

\shorttitle{OH in the Extreme Carbon Star IRC+10216}
\shortauthors{Ford et al.}

\begin{document}

\title{Detection of OH Towards the Extreme Carbon Star IRC+10216}
 
\author{K. E. Saavik Ford, David A. Neufeld}
\affil{Dept. of Physics \& Astronomy, The Johns Hopkins University, 
3400 N. Charles St., Baltimore, MD 21218-2686}
\email{saavik@pha.jhu.edu, neufeld@pha.jhu.edu}

\author{Paul F. Goldsmith,}
\affil{National Astronomy and Ionosphere Center, Department of 
Astronomy, Cornell University, 610 Space Sciences Building, 
Ithaca, NY 14853-6801}
\email{pfg@astrosun.tn.cornell.edu}

\and

\author{Gary J. Melnick}
\affil{Harvard-Smithsonian Center for
Astrophysics, 60 Garden Street, Cambridge, MA 02138}
\email{melnick@cfa.harvard.edu}

\begin{abstract}

We report the detection of the $1665$ and $1667\,$MHz 
main lines of OH (hydroxyl) and upper limits on the 
$1612\,$MHz satellite line of OH towards the carbon-rich AGB 
star, IRC+10216.  We find a beam averaged fractional abundance 
$x({\rm OH})\sim 4 \times 10^{-8}$.
This detection supports the identification by Melnick et al. (2001)
of the $1_{10}-1_{01}$ transition of water vapor with a
$556.936\,$GHz emission feature detected towards IRC+10216, since OH
is the expected photodissociation product of water vapor.
The shape of the OH lines, however, differs significantly
from the shape expected based on the observations of 
Melnick et al. (2001).  Possible explanations for the anomalous
shapes of the $1665$ and $1667\,$MHz lines are
discussed.  The most likely explanations for the unexpected
OH line shapes are either masing or an asymmetric distribution
of OH molecules around IRC+10216.

\end{abstract}


\section{Introduction}

IRC+10216 is the prototypical late-stage, carbon-rich asymptotic giant 
branch (AGB) star.  This star is extremely well-studied, 
due to its close proximity and large mass loss rate; the
star is losing mass at a rate of 
$\sim 3 \times 10^{-5}\,$M$_{\odot}\,{\rm yr^{-1}}$
(Glassgold 1996), producing a dense, 
dusty circumstellar envelope, well shielded
from interstellar ultraviolet (ISUV) radiation.  
The dense, shielded environment of IRC+10216's envelope
is home to a rich circumstellar chemistry.  To date, more than 50 molecules
have been detected around IRC+10216.
Due to the extreme carbon-rich nature
of the source (C/O $\gtrsim 1.4$) (Glassgold 1996), the detection of any
oxygen-bearing molecules other than CO and small amounts
of SiO and HCO$^+$ was entirely unexpected.  
Recently, however,
Melnick et al. (2001) reported the detection of an emission feature
at $556.936\,$GHz, attributed to the $1_{10}-1_{01}$ transition
of water vapor.  The detection of water vapor in such a
carbon-rich circumstellar envelope was interpreted by Melnick et al. (2001)
as evidence
for the existence of an extrasolar cometary system, analogous
to the Solar System's Kuiper Belt, in orbit around IRC+10216.
In this system, the luminosity of the
central star has increased dramatically due
to the later stages of 
post-main sequence stellar evolution, 
causing the icy bodies in the Kuiper Belt analog to 
vaporize (Ford \& Neufeld 2001)
and produce the water vapor observed by Melnick et al. (2001).

Of possible concern for the cometary hypothesis is the existence of 
a peculiar subclass of carbon stars which
exhibit OH and H$_2$O maser emission, as well as silicate emission
features (e.g. Engels 1994).  These features would
normally be unexpected in carbon stars, for the reasons
explained above.  This subclass also
shows a number of other features which distinguish
peculiar carbon stars from normal carbon stars, including
relatively high $^{13}$C/$^{12}$C ratios, IRAS colors compatible with
low mass loss rate oxygen stars, and depleted s-process element
abundances relative to other carbon stars.  
These peculiar carbon
stars may represent objects in the transition stage between oxygen
star and carbon star, and their photospheric chemistry is likely
to be out of thermochemical equilibrium, or they may be members of
binary pairs of one
oxygen star and one carbon star (Wallerstein \& Knapp 1998).  
In any case, IRC+10216 displays
none of the ancillary characteristics associated with these
peculiar carbon stars, aside from OH and H$_2$O emission.
Due to the large C/O ratio in IRC+10216 and the fact that
it is not a member of a binary system, as well as the
lack of ancillary characteristics, we exclude the
possibility that OH or H$_2$O emission from IRC+10216
arises from membership in this peculiar class of carbon stars.

Since the Melnick et al. (2001) result was based on the
detection of a single line of water vapor, we became interested
in finding a method of verifying the 
identification of water vapor with the emission feature at
$556.936\,$GHz.  To this end, we realized that the water vapor emitted
by icy bodies in the cometary system would be carried
by the circumstellar outflow and shielded by the dust, until the outflow
became diffuse enough to allow penetration by the ISUV field.
Once the ISUV field could penetrate the outflow, 
the water vapor would be photodissociated into OH and atomic
hydrogen.  The OH could be detected, and such a detection
would yield support for the Melnick et al. (2001) identification.  
Based on our model for the production of OH, we proposed a deep
search for OH around IRC+10216 (a previous search by 
Andersson, Wannier \& Lewis (1992) resulted in a non-detection); 
we were awarded $30$ nights at the Arecibo 
Observatory\footnote{The Arecibo Observatory is part of the 
National Astronomy and Ionosphere Center, which is 
operated by Cornell University under a cooperative 
agreement with the National Science Foundation.}
for our search.
We present our observations in \S 2, discuss
the implications of our results in \S 3 and 
present our conclusions in \S 4.

\section{Observations}

The radio telescope at Arecibo Observatory (AO) is an
altitude-azimuth telescope with a fixed $304.8\,$m primary 
spherical reflector 
and a movable Gregorian dome which contains secondary and
tertiary reflectors, as well as multiple receivers.  For the
observations discussed in this paper,
the L-wide receiver was used with dual linear polarizations
with each of four boards of the correlator in a 9-level mode,
centered on the frequency of interest 
and covering a bandwidth of $3.125\,$MHz divided
into $1024\,$channels, resulting in an unsmoothed velocity resolution
of $0.55\,{ \rm km\, s^{-1}}$ at $1665$ and $1667\,$MHz.
Observations were taken 2001 November 27 -- December 10; 
2002 February 2 -- 3, 12, 14, 23; 2002 March 2 -- 4, 24, 27 -- 31;
and 2002 April 1, which resulted in
a total on-source integration time of $1613\,$minutes.
At least one flux calibration was done during each observing session,
on one, two or three of the following sources: 
B0838+133, J0935+086, B1117+146, B1040+123 or B1005+077, 
depending upon the amount of time available.  Our calibration
sources were observed at a variety of zenith angles, so we
may take an average gain from these measurements as representative
of the telescope gain for our target observations.
We found the telescope gain to be
$8.2\,$K/Jy at $1666\,$MHz.
All observations were conducted using standard position switching,
with equal duration $5\,$minute on-source and off-source scans, except on 
2001 November 28,
when the scan duration varied between $3$ and $5\,$minutes in an
effort to maximize observing time.
The transitions of primary interest, the $1665$ and $1667\,$MHz main lines
of OH, were observed during all sessions.  Other transitions which were
only observed during some of the sessions
were the $1612$ and $1721\,$MHz satellite lines of OH, the
$1420\,$MHz hyperfine line of HI, 
the $1347\,$MHz $2_{-1}-2_1$ line of $\nu_2=1$ HCN and
the $1743\,$MHz $3-2$ line of HC$_9$N.

\begin{figure}
\plotone{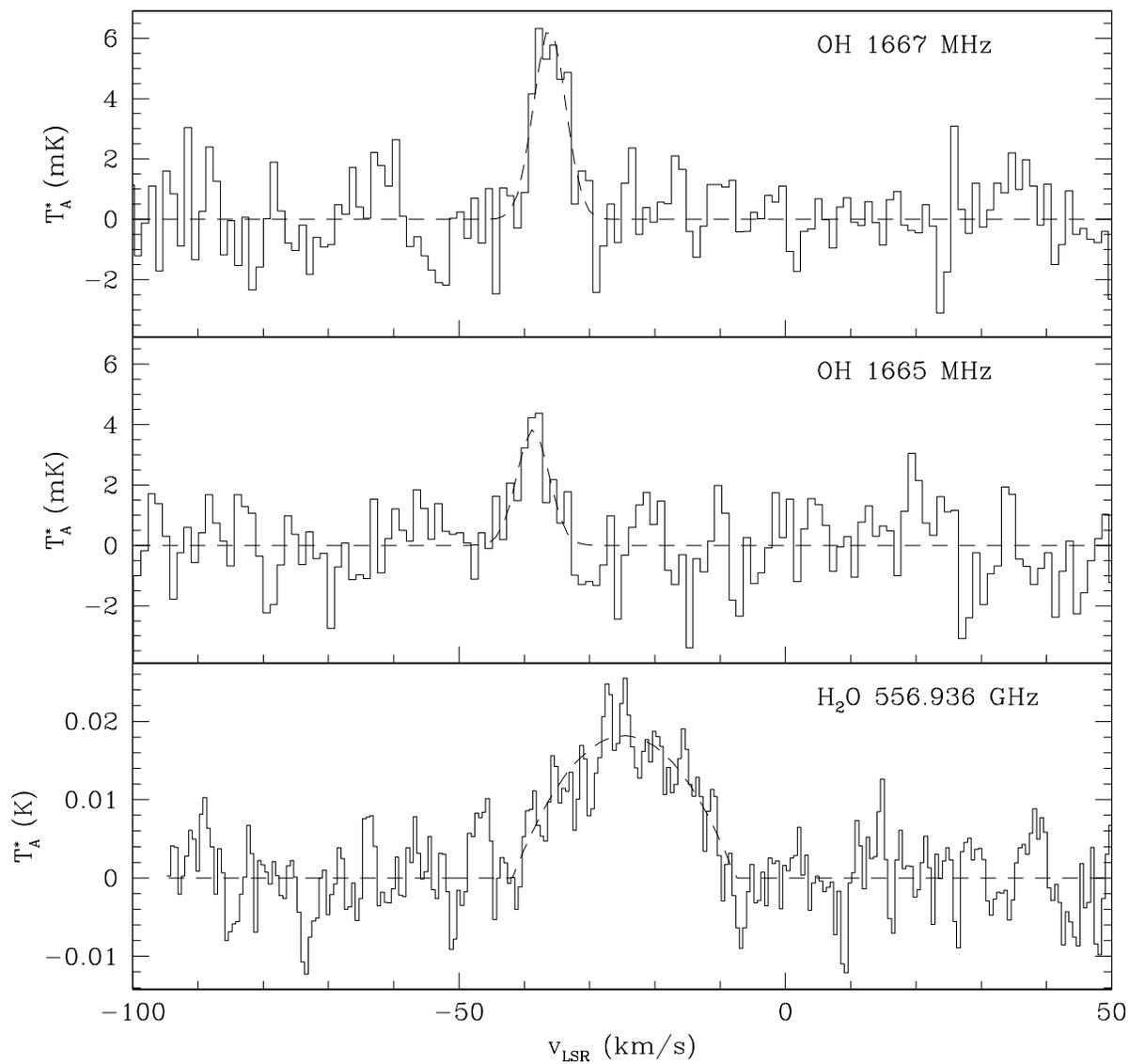}
\caption{Spectra of IRC+10216 at $556.936\,$GHz (Melnick et al. 2001), 
and $1665$ and $1667\,$MHz. For the $556.936\,$GHz line, the dashed
line is a fitted parabola, as expected for an optically thick expanding
spherical shell.  The dashed lines in the $1665$ and $1667\,$MHz spectra
are fitted Gaussians.  Parameters for all fits are 
listed in Table \ref{linefit}.
\label{spec}}
\end{figure}

Data reduction was accomplished by converting the raw AO data
to Continuum and Line Analysis Single-dish Software (CLASS) format.
To prevent contamination of the data by
RFI signals, each scan was examined by eye, and
any scans with apparent RFI (i.e. those with prominent, rapidly
varying signals) 
were eliminated.  The remaining scans
were summed to produce a total spectrum for each observed frequency.
The total spectrum was then smoothed, using the CLASS ``smooth''
macro, resulting in a smoothed 
velocity resolution of $1.1\,{\rm km\,s^{-1}}$.
A first order polynomial baseline was subtracted from the
smoothed data.
In the case of the OH main lines, an emission 
feature was readily apparent
in the final data, and a Gaussian profile was fit to each
feature.  The calibrated main line spectra, 
together with the Gaussian fits, are displayed in Figure \ref{spec}.
The fit parameters for these lines are listed in Table \ref{linefit}.
The errors listed in the table are $1\sigma$; thus the $1665$ and $1667\,$MHz
lines are detected at a significance of
$4.7\sigma$ and $8.8\sigma$, respectively.
We determine an average fractional abundance for OH
of $x({\rm OH})\sim 4 \times 10^{-8}$ using
the $1667\,$MHz line flux, $F_{1667}=39\,{\rm mK\, km\, s^{-1}}
=4.8\,{\rm mJy\, km\, s^{-1}}$, and the relationship
between $F_{1667}$ and the number of OH molecules, 
$N({\rm OH})=4 \pi D^2 F_{1667}/f_u A_{ul} h\nu$, where
$D=170\,$pc is the distance to IRC+10216, $f_u=5/16$ is the
fraction of molecules in the upper state (the value of 5/16
is plausible for the most likely excitation conditions),
$A_{ul}=7.8 \times 10^{-11}\,{\rm s^{-1}}$ is the Einstein A
coefficient for the transition and $h\nu$ is the 
energy of the transition.  Once we have found $N({\rm OH})$,
we may divide by the number of H nuclei in the Arecibo beam to
find $x({\rm OH})$.  Based on a density of H nuclei
of $n_H=3\times 10^{37}(\dot M_{-5}/v_{exp,6})/r^2\,$cm$^{-3}$
(Glassgold, 1996), where $\dot M_{-5}=3$ is the stellar mass
loss rate in units of $10^{-5}\,{\rm M_{\odot}\, yr^{-1}}$,
and $v_{exp,6}=1.4$ is the expansion velocity\footnote{The
expansion velocity determined by Melnick et al. (2001)
from the water vapor line in IRC+10216 is
$v_{exp,6}=1.7$.  We use $v_{exp,6}=1.4$ instead,
since this is the value found previously for IRC+10216
(e.g. Skinner et al. 1999),
and because there are large
errors associated with the Melnick et al. measurement.}
of the circumstellar envelope
in units of $10^6\,{\rm cm\, s^{-1}}$,
we find
$7.8 \times 10^{55}$ hydrogen nuclei, assuming a maximum radius
of $r_{max}=10^{17}\,$cm.  
We choose $r_{max}=10^{17}\,$cm
because we expect nearly all of the OH to lie inside of this radius,
and we do not correct for beam sensitivity since this is within
the half-power beamwidth of AO and any such correction would
be at the level of a few percent at most.
This yields a rough
determination of $x({\rm OH})$ which is relatively model
independent but somewhat misleading.  Since we believe the
OH to be in a photodissociated shell or ring around IRC+10216,
the OH fractional abundance in the outer regions of the circumstellar envelope
around
IRC+10216 may reach values several
times larger than our minimum.

We failed to detect any signal originating
in the envelope of IRC+10216 at the five other
frequencies we observed.
Severe
radio frequency interference (RFI) across all observations
prevents us from placing meaningful upper limits 
on the strength of the $1721\,$MHz satellite line of OH or the 
$1743\,$MHz line of HC$_9$N.
Strong contamination from
interstellar gas at the
velocity of IRC+10216
prevents us from making any strong statements about
stellar emission at $1420\,$MHz, the hyperfine line of HI.
Our results suggest the necessity of 
interferometric observations to detect
and map the contribution to HI emission from the star.
For the remaining lines, we were able to determine 
upper limits on the observed flux. 
These are also listed in Table \ref{linefit}.

\section{Discussion}

In conjunction with the observations discussed above,
we constructed several models which attempted to
predict the strength and shape of the
OH lines in IRC+10216.
First, we used a photodissociation model for H$_2$O and OH,
assuming a dusty spherical outflow, illuminated from 
the exterior by the interstellar
ultraviolet (ISUV) field.
We used photodissociation
rates and dust opacities based on Roberge et al. (1991).
The rates in Roberge et al. (1991) were calculated
for a plane-parallel slab, so a correction for
the spherical geometry of the circumstellar envelope around 
IRC+10216 was applied.  This
correction is described in detail in Appendix \ref{corr}.
Our model assumed spherical shielding of
the molecules due to the spherical distribution of
dust around IRC+10216, but did not assume
a spherical molecular distribution, since molecular 
self-shielding was neglected, as is appropriate under the condition
of small molecular abundances. A given model produces
the relative abundance of H$_2$O and OH as 
a function of radial distance.
We generated a number of
abundance distributions by varying the unshielded molecular 
photodissociation rate,
the stellar mass loss rate, $\dot M$, and the dust opacity.
The unshielded molecular 
photodissociation rate yields the strength of the ISUV field, 
the stellar mass loss rate determines the density of
the shielding dust, and the dust opacity indicates the
shielding strength of the dust.
Photodissociation models were parameterized 
by assuming a photodissociation
rate, $\Gamma_i$, of the form 
$\Gamma_i (A_v)=2 C_i G_0
\exp(-\alpha_i A_v - \beta_i A_v^2)\, {\rm s^{-1}}$, 
where $i$ represents either OH or H$_2$O, 
$A_v$ is the depth in visual magnitudes and $C_i$, $\alpha_i$ and $\beta_i$
are factors derived from Roberge et al. (1991) using the procedure
described in Appendix A.
The factor $G_0$ represents the relative strength
of the ISUV field and $G_0=1$ defines the average ISUV field
(Draine 1978).
We also considered variations in the ratio of $A_v$ to $N_H$, where
$N_H$ is the column density of hydrogen nuclei.  This ratio, $A_v/N_H$,
represents how much extinction, and therefore molecular shielding,
is produced per unit column density
of hydrogen nuclei.  For standard interstellar grains considered by
Roberge et al. (1991), 
$A_v/N_H=(A_v/N_H)_0\equiv 5.6\times 10^{-22}\, {\rm mag\, cm^{-2}}$.  
The photodissociation model was then used as an
input to a second program, which calculated the expected strength
and shape of the OH emission lines, including instrumental effects
(ie: convolution of beam sensitivity with target intensity)
and Gaussian microturbulence.  For the beam convolution, we assumed
the antenna response is a Gaussian, with 
$\theta_e=104^{\prime \prime}$
where $\theta_e$ is the angular scale on which the sensitivity
is reduced by a factor $e$ relative to its on-axis value.
This corresponds to a half-power beamwidth, ${\rm HPBW}=2.89^{\prime}$
(Heiles 1999).

\begin{figure}
\plotone{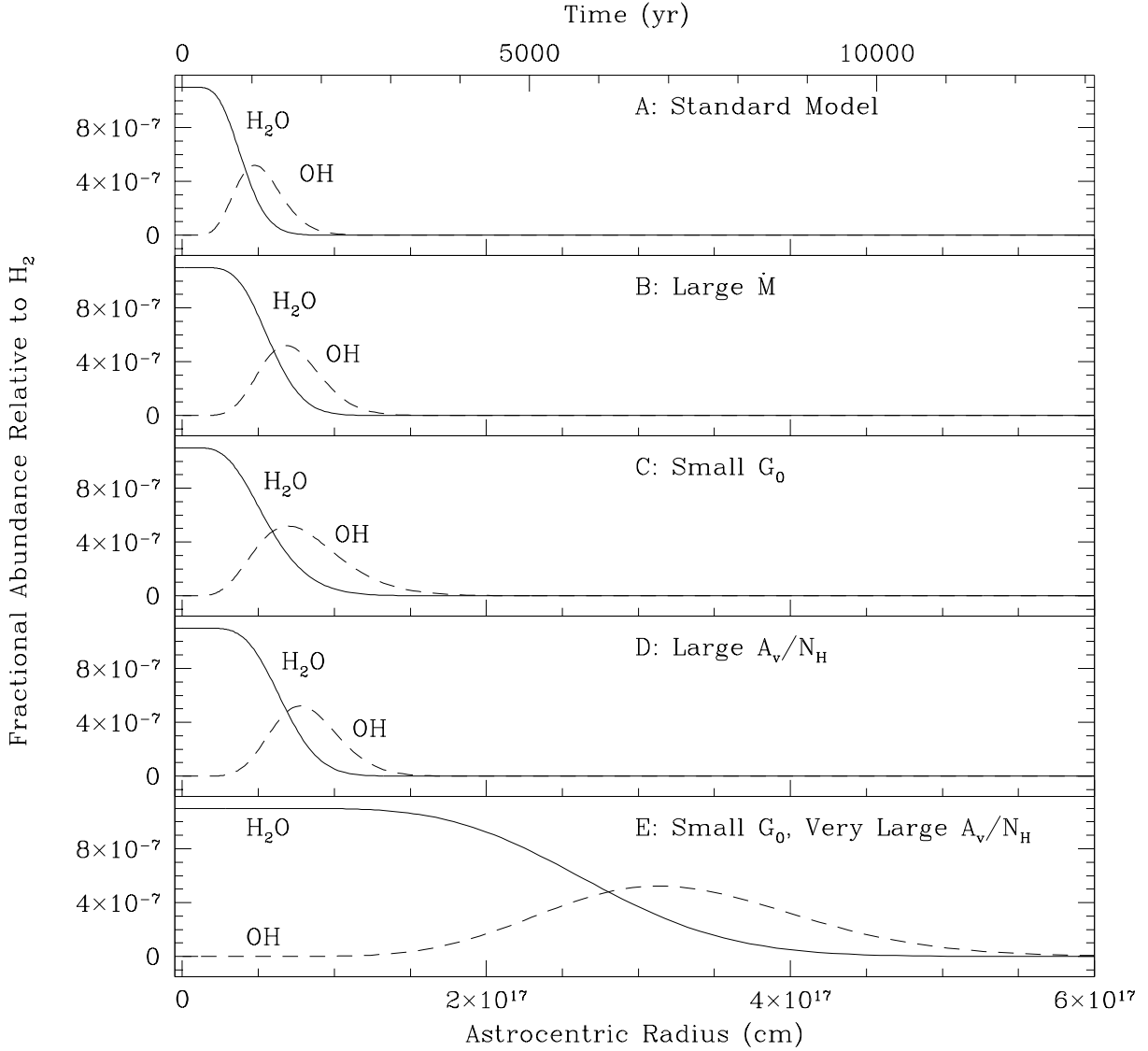}
\caption{Fractional water and OH abundance as a function of radial distance.  
Also plotted on the upper axis is the time for a parcel of gas
to travel a radial distance $R=v_{out}t$, based on a
terminal velocity for the circumstellar outflow of
$v_{out}=14.5\,{\rm km\,s^{-1}}$.
The solid line is the fractional water abundance and the dashed
line is the fractional OH abundance.  The fractional 
abundances are based on the
photodissociation models discussed in the text.  For input
parameters, see Table \ref{phottab}. \label{abund}}
\end{figure}

We have chosen five representative photodissociation models,
with parameters listed in Table \ref{phottab}, labeled by the
letters A through E.  Model A assumes the usual ISUV field strength,
$G_0=1$, normal interstellar grains, $A_v/N_H=(A_v/N_H)_0$, and a mass
loss rate, $\dot M=3 \times 10^{-5} {\rm M_{\odot} yr^{-1}}$, which
is in the middle of the range of mass loss rates measured for IRC+10216.
Models B through E demonstrate the effect of departures from the values
of Model A.  The results of our photodissociation code for 
Models A through E are plotted in Figure \ref{abund}.
The solid line represents the fractional abundance of H$_2$O relative to
molecular hydrogen, $x({\rm H_2O})$, while the dashed line is the
fractional abundance
of OH relative to molecular hydrogen, $x({\rm OH})$, both as a function
of astrocentric radius.  All models are normalized to a peak
fractional water abundance of $x({\rm H_2O})=1.1 \times 10^{-6}$,
consistent with the middle of the range of values for $x({\rm H_2O})$
determined by Melnick et al. (2001) from observations 
of the water vapor around IRC+10216.
Notice that for Models B through D, even relatively large
changes (factors of 2 or 3) in the 
values of $G_0$, $A_v/N_H$ and $\dot M$ have a relatively small
effect on the radial location of the peak in $x({\rm OH})$.  The small
value of $G_0$ in Model C and the large 
value of $A_v/N_H$ in Model D are
fairly unrealistic (Jura 1983), but the plots
of Models A through D demonstrate the range of possible sizes
of the OH shell with even marginally realistic parameters.
Indeed, the only photodissociation model which produces a
shell significantly larger than Model A is Model E, which
requires an implausibly large $A_v/N_H$ and a relatively
improbable $G_0$.

\begin{figure}
\plotone{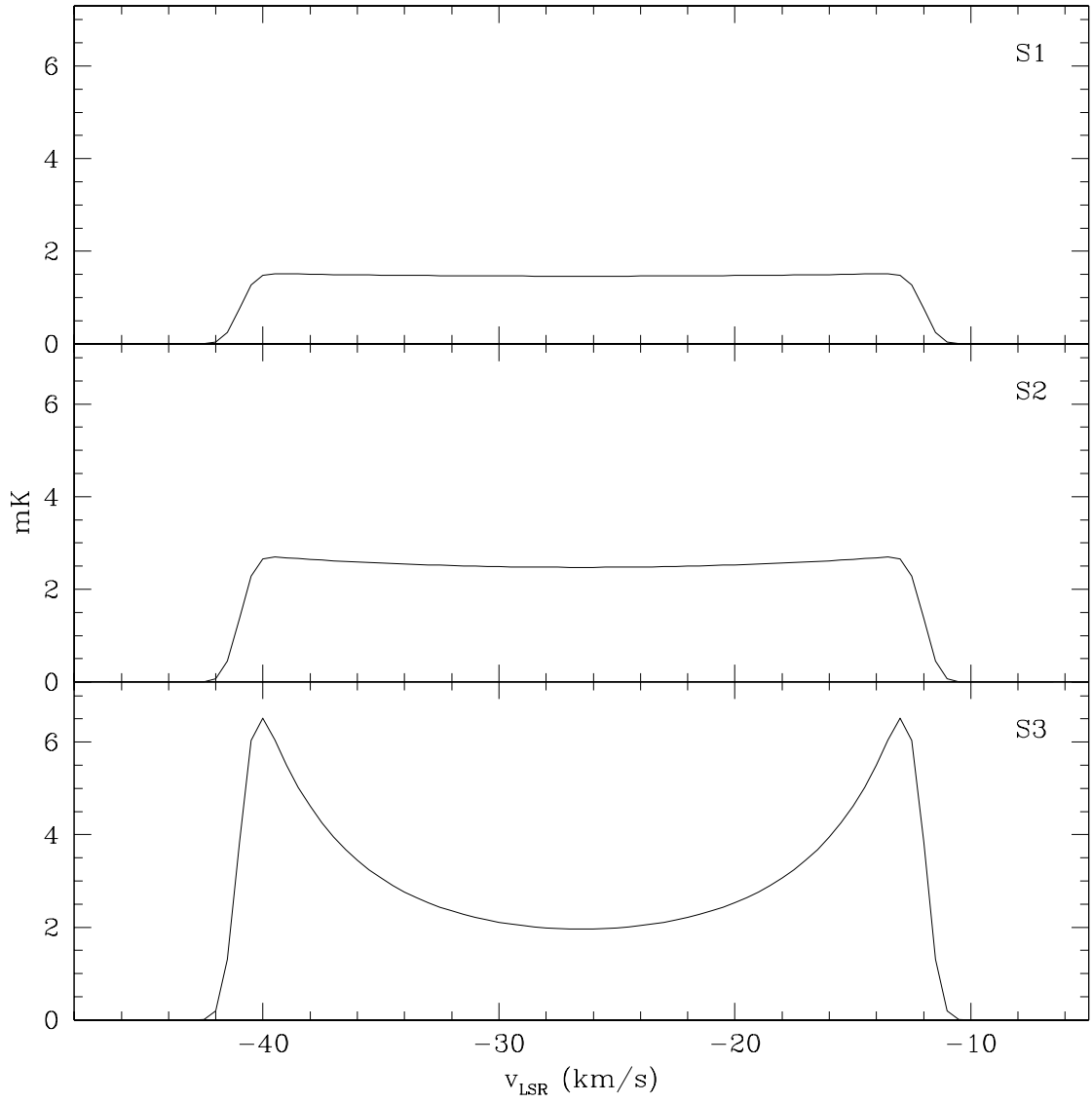}
\caption{Models of line profiles for spherical shells.
See Table \ref{linetab} for input parameters. \label{sprofile}}
\end{figure}

Initially, we consider the line shape produced by OH
distributed in a spherical shell.  Based on
photodissociation Model A, our line
profile program predicted that the OH emission would come from an almost
entirely unresolved spherical shell, and the line profile
would thus be very nearly a ``top hat'' profile. The predicted profile
is displayed in Figure \ref{sprofile} as line model S1,
and the parameters for that line model as well as several other
line models are listed in Table \ref{linetab}.  
After initial observations
indicated the absence of detectable
emission at $-26.5\,{\rm km\,s^{-1}}$, the systemic 
velocity of the source (Cernicharo, Gu\'elin \& Kahane 2000), 
other models
were discussed. One possibility we considered was that
the OH shell was larger than we originally predicted, 
due either to a larger than expected
dust opacity or to a weaker than expected ISUV field. If the OH shell
were sufficiently enlarged, it would be partially
resolved by AO and would produce a double-peaked line profile.
Unfortunately, as discussed above, extremely unreasonable parameters
are required to produce a substantially enlarged OH shell.
Even using photodissociation model C, displayed in 
Figure \ref{sprofile} as line model S2, the line profile remains
essentially a ``top hat''.  Only photodissociation Model E
produces a double peaked line (line model S3 in Figure \ref{sprofile}).

\begin{figure}
\plotone{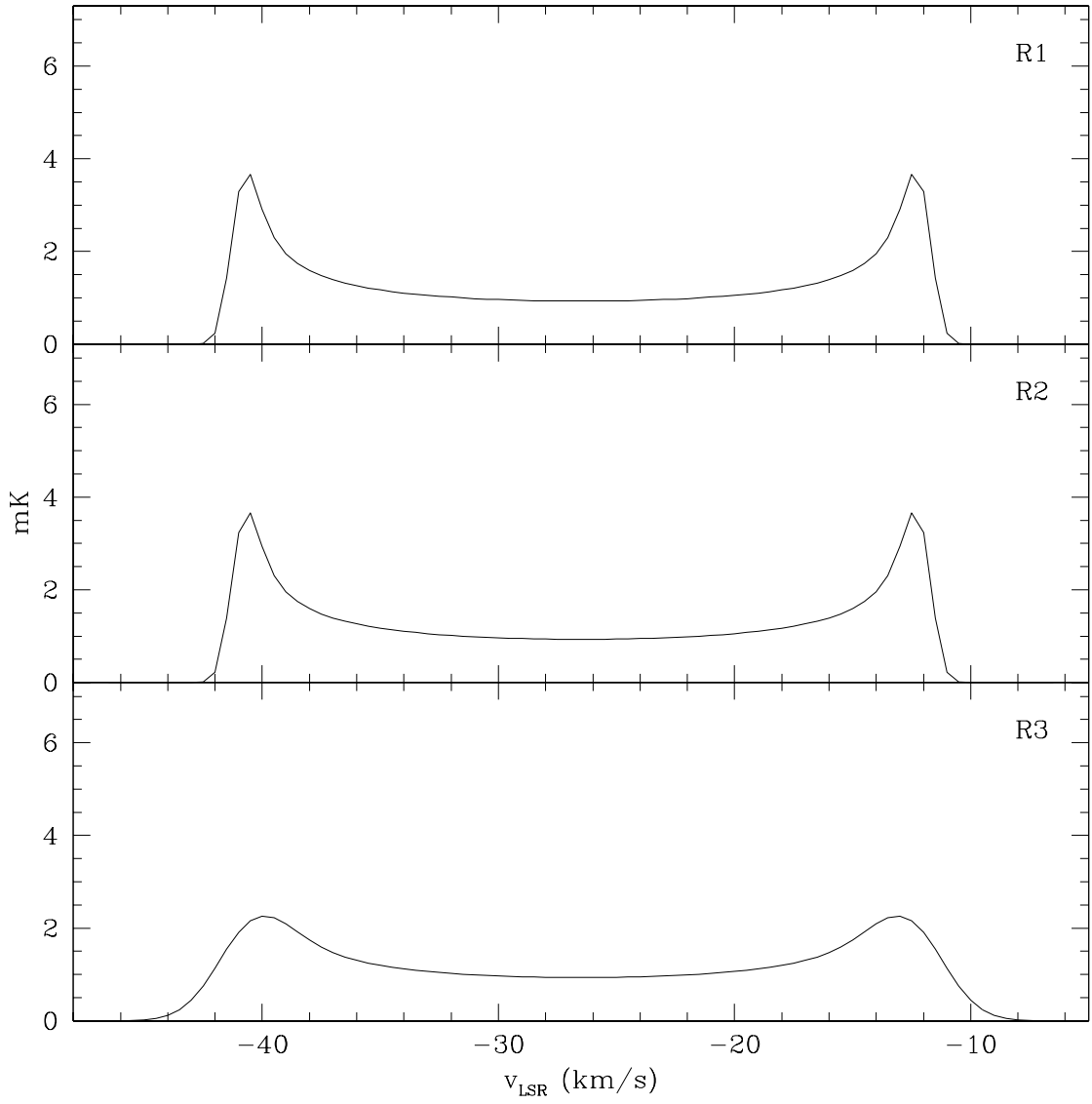}
\caption{Models of line profiles for rings. 
See Table \ref{linetab} for input parameters. \label{rprofile}}
\end{figure}

The final possibility we considered was an edge-on ring, which
would also give rise to a double-peaked profile
for the optically thin OH, even if the ring
were unresolved.  A ring structure for the OH would arise naturally
if the water vapor were distributed in a disk, as might
be expected for a cometary system analogous to the Kuiper Belt.
Several ring
profiles are plotted in Figure \ref{rprofile}; 
the input parameters for the ring line models are listed in
Table \ref{linetab} together with the spherical line models.
The rings are defined in spherical coordinates as extending from
$\theta={\pi \over {2}}-{\psi \over {2}}$ to 
$\theta={\pi \over {2}}+{\psi \over {2}}$, where $\psi$ is the
opening angle of the disk/ring, taken as an adjustable parameter in our model; 
$\psi=180\,$degrees is a sphere.
Note that changing the opening angle from
$10\,$degrees (model R1) to $1\,$degree (model R2)
has little effect on the line shape.  
We note
that an unresolved expanding disk would lead to an elliptical profile for
the optically thick water line.  The water line profile 
obtained by Melnick et al. (2001)
is plotted in Figure \ref{spec}. It does not have a sufficiently large
signal to noise ratio to distinguish between an elliptical profile and
the parabolic profile expected for an expanding sphere.
Finally, in model R3, we
demonstrate the effect of changing the
assumed microturbulent velocity, $b$, from the value measured
previously for several other lines in IRC+10216, 
$b=0.65\,{\rm km\, s^{-1}}$ (Skinner et al. 1995),
to $b=2.0\,{\rm km\, s^{-1}}$.  Such a change increases
the width of the peaks and tends to decrease
the ratio of height of the line peak to the height of the
line center.  The microturbulent velocity of
the OH gas could be increased by the injection of energy due to 
the photodissociation process itself.

Although the predicted line shape varies substantially 
depending on the parameters of the photodissociation 
model and geometry, the predicted line strength is 
primarily a function of the assumed fractional water abundance.  
For a given photodissociation model and geometry, the 
line strength varies linearly with assumed fractional water abundance.  
Unfortunately the fractional water abundance is uncertain by more than 
a factor of two, with possible values in the range 
$x({\rm H_2O})=4-24 \times 10^{-7}$ and a preferred value
of $x({\rm H_2O})=1.1 \times 10^{-6}$
(Melnick et al. 2001).  The large range of possible 
values for the fractional water abundance is due primarily to the 
uncertainty in the distance to IRC+10216, which is 
unknown to a factor of $\sim 2$.  Various authors report
reasonable distance estimates between $300\,$pc (Doty \& Leung 1997) and
$110\,$pc (Groenewegen, Van Der Veen \& Matthews 1998).  
We adopt a distance of $170\,$pc (e.g. Winters, Dominik \& Sedlmayr 1994; 
LeBertre 1997; Skinner et al. 1999) 
throughout.
For line model S1 (based on photodissociation model A) and
$x({\rm H_2O})=1.1 \times 10^{-6}$, we predict a line strength of
$10.5\,{\rm mJy\,km\,s^{-1}}$ (or $86.1\,{\rm mK\,km\,s^{-1}}$) 
for the $1667\,$MHz line, 
However, due to the
uncertainty in the fractional water abundance, this corresponds to
possible line strengths ranging from $3.82$ to $22.9\,{\rm mJy\,km\,s^{-1}}$
($31.3$ to $188\,{\rm mK\,km\,s^{-1}}$).  The predicted 
line fluxes listed in Table \ref{linetab}
are for the $1667\,$MHz line and are 
calculated assuming $x({\rm H_2O})=1.1 \times 10^{-6}$.
The $1665\,$MHz line is $5/9$ the strength of
the $1667\,$MHz line in LTE for optically thin emission.

While the observed strengths of the $1667$ and $1665\,$MHz lines 
agree well with our line strength predictions,
the observed OH line shapes are in very poor agreement with 
our predictions. 
The observed line profiles are neither ``top hats'' nor double-peaked;
indeed, they are not even symmetric about $v_{LSR}=-26.5\,{\rm km\, s^{-1}}$, 
the systemic
velocity of IRC+10216.
This asymmetry is unique\footnote{Observations by Lucas \& Cernicharo (1989)
and Lucas \& Guilloteau (1992) found asymmetric line profiles for HCN
in IRC+10216.
However, this emission originates in the inner regions of the
circumstellar outflow (inside the acceleration zone).  The observed
asymmetry was attributed to masing.}
 - every other molecule
observed in the outer envelope of IRC+10216 has a profile which is
symmetric about the systemic velocity (e.g. Cernicharo et al. 2000).
Rather, the $1667$ and $1665\,$MHz emission features
are blueshifted with respect
to the systemic velocity, and they
have central velocities which differ slightly
(see Table \ref{linefit}).  Based on the errors quoted in 
Table \ref{linefit}, we might conclude that the velocities of the
two features differ at a statistically significant level.  However,
we note that the errors are based on fitting a Gaussian profile to 
a line which is not likely to be a true Gaussian. A more appropriate
(and conservative)
estimate of the error associated with the line center determination is 
simply the velocity resolution of our spectrum.  For the 
$1665$ and $1667\,$MHz lines, the size of a resolution element after
smoothing is 
$1.1\,{\rm km\,s^{-1}}$.  Given this error estimate, we conclude that
the $1665$ and $1667\,$MHz lines may have the same central velocity.

\begin{figure}
\plotone{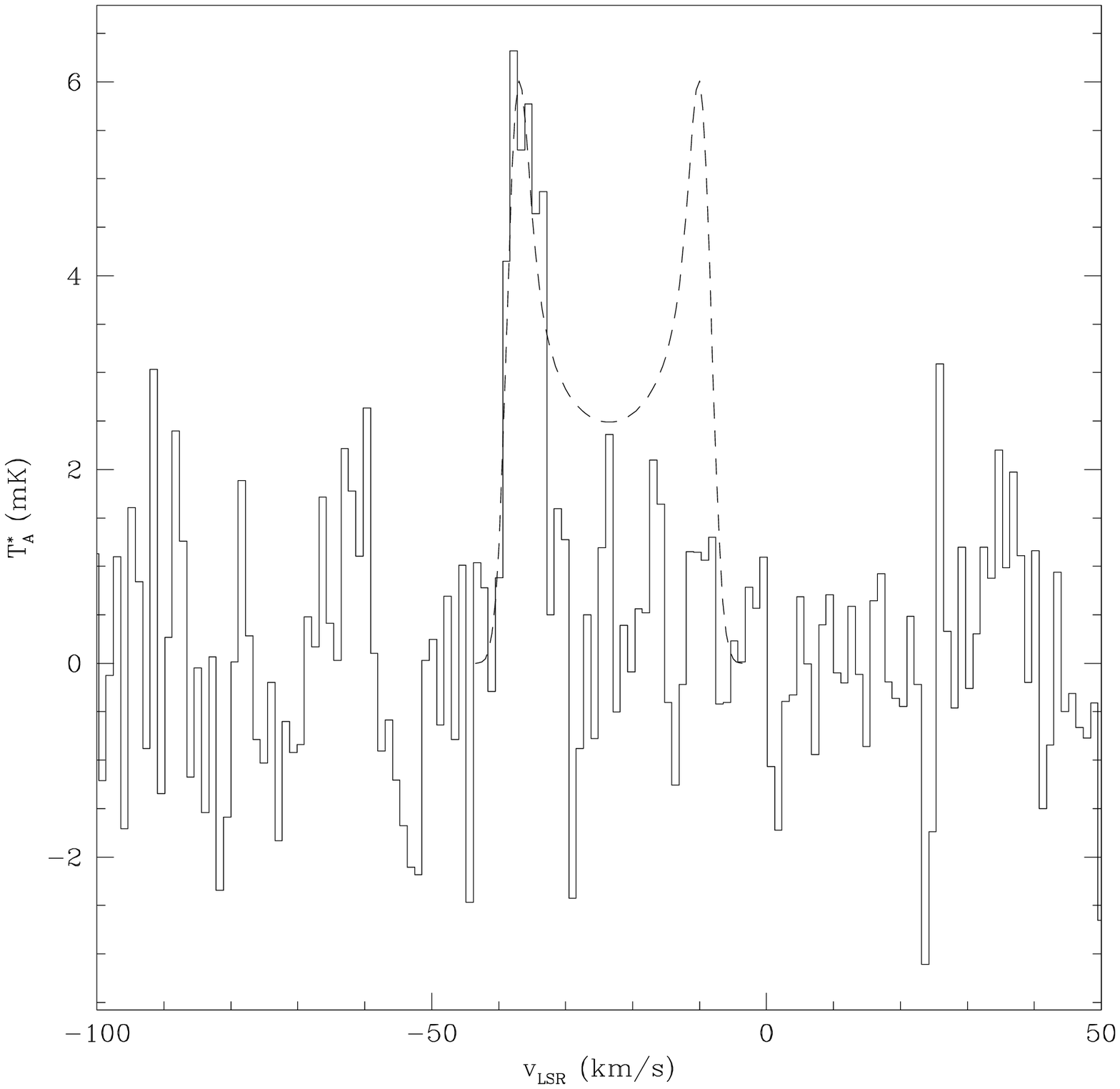}
\caption{1667 spectrum with best fit model overplotted. 
Model R3 (see Table \ref{linefit}) provides the best single
peak fit.  In order to match the velocity of the observed line
with the blueshifted peak of the model, we assume a 
$v_{LSR}=-23.5\,{\rm km\, s^{-1}}$,
which represents a 
redshift of $3\,{\rm km\, s^{-1}}$ with
respect to prior measurements of the $v_{LSR}$ of IRC+10216 
(e.g. Cernicharo et al. 2000).  
The missing redshifted peak may be explained by
a spatially asymmetric distribution of OH 
or maser amplification of the blueshifted peak (see text for details).
\label{datamod}}
\end{figure}

For comparison with our pre-observation models, we plot
our ``best fit'' model over the data for the $1667\,$MHz line
in Figure \ref{datamod}.
Our ``best fit'' is model R3, except that we must assume a fractional water 
abundance of $2.4 \times 10^{-6}$ in order to match the line fluxes
for velocities between $-40$ and $-30\,{\rm km\,s^{-1}}$.
Also, in order to match the velocity of the observed line
with the blueshifted peak of the model, we assume a 
$v_{LSR}=-23.5\,{\rm km\, s^{-1}}$,
which represents a
redshift of $3\,{\rm km\, s^{-1}}$ with
respect to prior measurements of the $v_{LSR}$ using other species
(e.g. Cernicharo et al. 2000). 
It is obvious from the plot that
the blueshifted peak of the model fits the data reasonably well, but the
redshifted peak of the model is entirely missing from the data.

The observed shapes of the $1667$ and $1665\,$MHz OH lines
are quite unexpected.  There are only three general mechanisms
which might create the observed line profiles;
these mechanisms are: 1) absorption, 2) masing or 
3) an intrinsic asymmetry in the spatial distribution of the
OH.  In the following sections, we examine each mechanism in
detail and discuss its plausibility.

\subsection{Absorption}
It seems reasonable to begin with the assumption that the OH 
emission from IRC+10216 is
intrinsically symmetric. If that were the case, then
in order to explain the OH line shapes, there 
must exist some absorbing
medium within the circumstellar envelope of IRC+10216 
which prevents emission from
the far (redshifted) side of the envelope from reaching us,
while emission from the near (blueshifted) side escapes and
is detected.  Since these observations were done at wavelengths
which are large, relative to previous observations of this star,
there is one source of opacity that could have previously gone
undetected - electron free-free absorption.  However, this 
absorption would be accompanied by thermal continuum emission.
In order to be an effective absorber of the redshifted half
of the OH line, the absorber would need to be optically thick
at a radius of $\sim 10^{17}\,$cm, where the OH distribution
peaks (though the peak radius of the OH distribution varies somewhat depending
on the parameters of the photodissociation model, $10^{17}\,$cm
is a reasonable value for most models).  At such a large radius, the
envelope gas temperature is quite cold, about $11\,$K,
and the thermal emission from the absorber would correspond
to a blackbody at the same temperature.
Unfortunately, our observations are 
not well suited to determining the
continuum brightness at $1665$ and $1667\,$MHz, but
Sahai, Claussen, \& Masson (1989)
have observed the continuum brightness 
at slightly higher frequencies, specifically at $15$ and $20\,$GHz.
The predicted flux densities due to free-free emission 
at $15$ and $20\,$GHz are $250$ and $140\,$mJy,
respectively, 
for an object which
is optically thick at $1665$ and $1667\,$MHz.
Sahai et al. (1989) find the flux densities to be only $6$ and $1.4\,$mJy
at $15$ and $20\,$GHz, respectively.
The absence of strong continuum emission indicates 
that at these frequencies, the envelope of IRC+10216 cannot be
optically thick at radii large compared to that of the 
AO beam, effectively ruling out the absorption scenario.
In addition, the electron density required to make the
envelope of IRC+10216 optically thick would be substantially
in excess of the densities presently expected based on
theoretical models of the envelope chemistry.

\subsection{Masing}
\label{masing}
A second possible explanation for the observed line profiles
is maser amplification of the $1667$ and $1665\,$MHz lines.
This possibility is attractive, because it is the only one
which could potentially be responsible for a velocity difference between
the $1667$ and $1665\,$MHz lines.  While the modeling of
circumstellar maser emission is difficult and a detailed model
of possible OH masers around IRC+10216 is beyond the scope of this paper,
we can compare our observations to other asymptotic giant
branch (AGB) stars with known OH maser emission.  
However, we should be cautious
about drawing strong conclusions based 
on this comparison, due to the fact
that all stars with confirmed circumstellar OH masers 
are oxygen-rich stars with much higher fractional OH abundances or
carbon-rich stars which have recently made the transition
from the oxygen-rich state and also have higher fractional OH abundances than
IRC+10216.

There is at least one star which is known to have exhibited OH
emission with velocity characteristics similar to IRC+10216, 
albeit at much larger flux densities.  In observations
reported by Etoka et al. (2001), the star R Crt 
displayed maser emission in
the $1667$ and $1665\,$MHz main lines, 
but not in the $1612\,$MHz satellite line; this in itself is not
particularly unusual among stars which exhibit circumstellar
OH masers.  More importantly, in observations on
1990 November 17, the $1667$ and $1665\,$MHz emission
appeared only at velocities blueshifted with respect to the
systemic velocity of R Crt, but with velocities which
were offset from each other by $\sim 1\,{\rm km\,s^{-1}}$.  Significantly,
the ratio of the $1667$ to $1665\,$MHz line strengths was $\sim 1.7$
during this observation, not far from the ratio of $1.8$ expected
in thermal equilibrium.  As the OH emission from R Crt is
known to be maser amplified, this example would seem to argue
for at least the possibility of masing in IRC+10216.

Several objections may be raised against the masing scenario, however, which
would need to be resolved before masing could be accepted as the source
of the line profile asymmetry in IRC+10216.
The first objection is that the predicted fluxes for IRC+10216
of the $1667$ and $1665\,$MHz lines in LTE from line model S1
are $1.3$ and $0.73\,{\rm mJy\, km\,s^{-1}}$ respectively,
for velocities between $-40$ and $-30\,{\rm km\,s^{-1}}$
and an $x({\rm H_2O})=4 \times 10^{-7}$.
These predictions are only a factor of $\sim 4$ less
than the observed
line strengths, even assuming the minimum possible $x({\rm H_2O})$,
and using line model S1, which predicts the smallest
flux density for velocities between $-40$ and $-30\,{\rm km\,s^{-1}}$.
The substantial blue/red
asymmetry of the line flux 
requires strong masing.  The asymmetry of
the $1667\,$MHz line
can be quantified by fitting a Gaussian with 
fixed parameters to the redshifted side of
the line and taking a ratio of the $3\sigma$
upper limit on redshifted flux, $F_{red}$,
with the measured flux from the blueshifted
side of the line, $F_{blue}=39.0\,{\rm mK\, km\, s^{-1}}$.  
For a Gaussian with a line center fixed at
$v_{LSR}=-10.9\,{\rm km\, s^{-1}}$ and a fixed width
of $5.8\,{\rm km\, s^{-1}}$, we find a flux of 
$4.38\pm11.4\, {\rm mK\, km\, s^{-1}}$, $3\sigma$.
We chose the center of the red feature so that the blue
and red sides of the line would be symmetric
about $v_{LSR}=-23.5\,{\rm km\, s^{-1}}$.
Thus, we find a $3\sigma$ upper limit
$F_{red}/F_{blue} \le 0.4$, or an amplification
of the blue side of the line by a factor of at least
$2.5$, clearly requiring significant amplification.
If the OH emission is substantially amplified
by maser action, the $1667$ and $1665\,$MHz lines ought
to have a brightness temperature comparable to or larger
than the temperature of the microwave background ($\sim3\,$K, or
about $3700\,{\rm mJy\,km\,s^{-1}}$ between 
$-40$ and $-30\,{\rm km\,s^{-1}}$).
This might lead one to conclude, based on our measurements, that
the observed $1667$ and $1665\,$MHz lines are too dim to be strongly
maser amplified.
Although the OH emission is not that bright, if maser amplification
were confined to a small spot, the beam dilution in AO's large
field of view could mean that our observations are still consistent
with amplification.  Specifically, the maser spot would need to
have a diameter smaller than 
$10\,$arcsec or $1700\,$AU at a distance
of $170\,$pc, assuming an intrinsic brightness of $3\,$K or larger.
This is not particularly small, considering that OH maser spots around
other evolved stars can have diameters smaller than 
$25\,$AU (Chapman, Cohen \&
Saikia 1991), so the dim lines do not exclude masing.

A far stronger objection stems from theoretical models
of circumstellar OH masers by Elitzur (1978).
Elitzur found that OH masers in oxygen-rich stars were
created by pumping from the far-IR emission of warm dust, 
and that for the $1667$ and $1665\,$MHz main lines, the maser
gain is proportional to the density of OH molecules (below a
threshold density).  Although Elitzur (1978) did not model
OH densities or gas temperatures 
as low as those expected in IRC+10216, a reasonable
extrapolation of his models indicates that $\vert \tau \vert$
would be $\ll 1$, where $\tau$ is the line center optical depth.

\subsection{Intrinsic Asymmetry}

The only remaining general class of explanations for
the OH line asymmetry is an intrinsic
asymmetry in the spatial distribution of the OH around IRC+10216.
There are two
possible mechanisms for producing such a spatially asymmetric
distribution.
One mechanism is an asymmetric ISUV field 
in the vicinity of IRC+10216; this would produce a larger fractional 
abundance of
photodissociation products on one side of the
circumstellar envelope relative to the other.  Since OH is
a photodissociation product, this scenario could produce
an asymmetric line profile if the viewing geometry were properly 
aligned.  However, all other molecular
photodissociation products would then have the same line
profile as OH.  This is not observed (e.g. Cernicharo et al. 2000).

The other way to produce an asymmetric distribution of OH
would be an asymmetric distribution of its parent molecule,
H$_2$O.  The observed water vapor emission is essentially 
symmetric
about the systemic velocity of IRC+10216.  However, the
H$_2$O emission and the OH emission do not arise
from precisely identical spatial regions (see Figure \ref{abund}).  
The time required for a parcel of gas to travel a
given radial distance can be found from the simple
relationship $R=v_{out}t$, where $v_{out}$ is the terminal
velocity of the stellar outflow.
Due to the direct relationship between the radial coordinate
and time ($R=v_{out}t$), the fact that the OH and H$_2$O
are not spatially coincident means that the present distributions
of OH and H$_2$O were formed at two different epochs, and
the OH and H$_2$O
need not have identical geometries at the present time.
If the spatial distribution of water
were asymmetric a few hundred years ago, but is symmetric now,
the OH and H$_2$O observations could be reconciled with the
intrinsic asymmetry scenario.

One possible explanation for
a change in the distribution of water vapor over such a relatively
short timescale involves the time dependence of the stellar mass
loss rate.  The circumstellar envelope of IRC+10216
contains multiple dust shells which are the result of episodic
mass loss.  Detailed images of these shells by Mauron \& Huggins
(1999, 2000), show that the shells are incomplete - they are
arcs rather than rings.  Additionally, the dust distribution
in the innermost regions of the envelope ($r\lesssim 15^{\prime \prime}$)
is an extremely asymmetric bipolar outflow.  The timescale on which
these dust shells are produced is of order a few hundred years.
A new mass loss study by Fong, Meixner \& Shah (2003) demonstrates
that mass loss in IRC+10216 has been ongoing for at least $7000\,$years
and that mass loss occurred in a clumpy and asymmetric fashion.
Rapid variations in the mass loss rate are also possible;
recent high-resolution IR images show dust distribution asymmetries on size 
scales as small as tens of milliarcseconds and imply mass loss
variations on timescales of years (Tuthill et al. 2000; Weigelt et al. 2002;
Men'shchikov, Hofmann \& Weigelt 2002).
Ford \& Neufeld (2001) showed that
the vaporization of water from icy bodies orbiting IRC+10216 is
sensitively dependent on the stellar flux impinging on an icy body.
Since the dust distribution is highly asymmetric, the luminosity
of the central star will ``leak'' out in an asymmetric and probably
quite patchy pattern, after multiple scatterings and dust absorption
and re-emission.
Thus, the flux impinging on different icy bodies at the same astrocentric
radius will not be necessarily be the same; it will depend on the
particular geometry of radiation leakage through the circumstellar dust.
Therefore, the water vaporization rate from the surface of a particular
icy body (and the distribution of water vapor in the circumstellar
outflow) will depend very substantially on the time varying 
dust distribution in the
circumstellar outflow.
Note that this scenario 
is still capable of producing the water line profile 
observed by Melnick et al. (2001).
If the water vapor is more or less 
off to one side of the star (from a terrestrial observer's 
perspective), there will be roughly equal amounts of 
material coming toward us and moving away from us, 
still yielding a roughly symmetric profile.
Additionally,
since the water line is optically thick, it is probable that
some clumpiness or patchiness in the water distribution would be smoothed
out in the line profile.

\section{Summary}

We have detected the $1665$ and $1667\,$MHz lines
of OH in emission
towards the carbon-rich AGB star IRC+10216. This detection
supports the identification of the $556.936\,$GHz emission
detected toward IRC+10216 by Melnick et al. (2001) with
the $1_{10}-1_{01}$ transition of water vapor.  The 
$1665$ and $1667\,$MHz line profiles are unexpectedly narrow
and blueshifted with respect to the systemic velocity
of IRC+10216.  We believe that the unanticipated line shape is
the result of either maser amplification or some intrinsic
spatial asymmetry in the OH distribution around IRC+10216, or
possibly both.  While we favor the spatial asymmetry explanation,
due to early theoretical work by Elitzur (1978),
further modeling of the excitation conditions of OH around
IRC+10216 using a code which includes strong masing
would help clarify the possible role of maser amplification
in this source.  Additional theoretical work may also be required
to understand the potential spatial asymmetry of OH, despite
the apparent symmetry of its parent molecule, H$_2$O.  Finally, while it would
be most desirable to spatially resolve the OH emission in IRC+10216,
such observations cannot presently
be completed within an acceptable
amount of telescope time.

\acknowledgments
We are grateful to the staff at AO, particularly 
Chris Salter, Phil Perillat, Karen O'Neil and
Mike Nolan.  We are especially grateful for the 
outstanding service observing work performed by Hector Hernandez.
We would also like to
thank Jennifer Wiseman, David Hollenbach, Al Glassgold and 
Barry McKernan for helpful discussions.
This work was supported by subcontract SV252005 
from the Smithsonian Institution; K.E.S.F. was also supported by
an American Dissertation Fellowship from the American Association
of University Women (AAUW).
The National Astronomy and Ionosphere Center is operated by Cornell
University under a cooperative agreement with the National Science Foundation.

\appendix
\section{Correction to Photodissociation Rates due to Spherical Symmetry}
\label{corr}

For the purposes of work described in this paper, 
we want a simple expression
for the photodissociation rates of H$_2$O and OH as a
function of depth, measured in magnitudes of visual extinction,
$A_v$, in a sphere which is illuminated
isotropically from the exterior by the ISUV field.
Calculations done by Roberge et al. (1991) (R91 hereafter) provide
rates for a plane-parallel slab geometry.  We describe here
a method which can be applied to the rates calculated
with plane-parallel
geometry which will give approximate rates for the spherical
case.

R91 calculated photodissociation and
photoionization rates for a number of species, as a
function of visual extinction into plane-parallel clouds
illuminated isotropically from both sides by the average ISUV
field (Draine 1978).  In general, the photodestruction rate, $\Gamma$, 
for a species, $i$, is
\begin{equation}
\Gamma_i (A_v)=4\pi 
\int_{\lambda_H}^{\lambda_i} J_{\lambda}(A_v) \sigma_i(\lambda) d\lambda.
\end{equation}
The rate, $\Gamma_i$, depends on the extinction, $A_v$, measured
inward from the surface of the cloud in units of visual magnitudes.
Here $\sigma_i(\lambda)$ is the wavelength dependent cross section
for the process of interest, and the integration runs from
$\lambda_H=91.2\,$nm to the threshold wavelength $\lambda_i$ 
for the process of interest. The quantity $J_{\lambda}(A_v)$
is the mean intensity of radiation at depth $A_v$ in 
photons cm$^{-2}$ s$^{-1}$ sr$^{-1}$ nm$^{-1}$.  R91
used the spherical harmonics method
to solve numerically the radiative transfer equation for
$J_{\lambda}(A_v)$ for clouds of $A_v^{tot}=1,10$ and $100\,$mag.
The solution for $J_{\lambda}(A_v)$ allowed R91 to
calculate the photodestruction rates for a large number of species
at various depths $A_v$ into each of their model clouds 
($A_v^{tot}=1,10$ and $100\,$mag).  In addition, R91
found that a bi-exponential function, of the form
\begin{equation}
\Gamma_i (A_v)=C_i \exp(-\alpha_i A_v - \beta_i A_v^2)\; {\rm s^{-1}}
\end{equation}
provided a reasonable fit to their numerical results, where $C_i$ is
the unattenuated photodestruction rate and $\alpha_i$ and $\beta_i$
are fit parameters, for a given species $i$.

A point $P$ in the interior of an R91 plane-parallel slab
is illuminated by radiation coming from many directions.
If we consider a slab with $A_v=100$, we can
assume that near one surface of the slab, essentially
all photodissociating radiation originates at the nearer
surface.  Under such conditions, 
the point $P$ ``sees'' rays diminished by an
``effective visual extinction'', $A_v^{\prime} =A_v/\cos \theta$
coming from a direction $\theta$, where the angle is measured
from the normal to the slab and $A_v$ is the depth measured
from the surface of the cloud to the point $P$ along the normal
ray.
In order to calculate photodissociation 
rates valid for a point $Q$ interior to a sphere, we realized 
that we should decompose the photodissociating radiation
arriving at point $P$ inside a slab into
constituent rays coming from a direction $\theta$.  This would
give us the photodissociation rate due to a single ray. Then,
we can use the result for a single ray to integrate over all
rays arriving at the point $Q$ in the
interior of the sphere.  Such a point
$Q$ ``sees'' rays diminished by a different ``effective visual extinction'',
$A_v^{\prime\prime}=\theta A_v/\sin \theta$.  We can then integrate over
all rays arriving at point $Q$, to determine the total photodissociation
rate for a point internal to a sphere.

So, given a formula for the photodissociation rate 
$\Gamma_i^{pp}=C_i \exp(-\alpha_i^{pp} A_v - \beta_i^{pp} A_v^2)$ in a
plane-parallel slab, we should be able to find
best fit values for the parameters
$\alpha_i^{r}$ and $\beta_i^{r}$ subject to the condition that
\begin{equation}
{\mid \Gamma_i^{pp} - \Gamma_i^{pr} \mid \over \Gamma_i^{pp}} \le 0.2
\end{equation}
for all $A_v$ of interest, where
\begin{equation}
\Gamma_i^{pr}=C_i \int_{0}^{\pi/2} 
\exp\left(-\alpha_i^{r} A_v^{\prime} 
-\beta_i^{r} \left(A_v^{\prime}\right)^2 \right) 
\sin \theta d\theta.
\end{equation}
Here $\Gamma_i^{pr}$ is the best fit photodissociation rate
at a point $P$ interior to a plane-parallel slab, found by
integrating over each individual ray terminating at point $P$.
The parameters $\alpha_i^{r}$ and $\beta_i^{r}$ are simply
the attenuation parameters along any single ray.
Once we have $\alpha_i^{r}$ and $\beta_i^{r}$, we can
repeat a similar procedure to find a best fit $\alpha_i^{sp}$ 
and $\beta_i^{sp}$.  Specifically, we wish to satisfy the
condition
\begin{equation}
{\mid \Gamma_i^{sr} - \Gamma_i^{sp} \mid \over \Gamma_i^{sr}} \le 0.2
\end{equation}
for all $A_v$ of interest, where
\begin{equation}
\Gamma_i^{sr}=C_i \int_{0}^{\pi} 
\exp\left(-\alpha_i^{r} A_v^{\prime \prime} 
-\beta_i^{r} \left(A_v^{\prime \prime} \right)^2 \right)
\sin \theta d\theta
\end{equation}
and
\begin{equation}
\Gamma_i^{sp}=2C_i \exp(-\alpha_i^{sp} A_v - \beta_i^{sp} A_v^2).
\end{equation}
Here $\Gamma_i^{sr}$ is the photodissociation rate
at a point $Q$ interior to a sphere, which is found by
integrating over all single rays arriving at point $Q$
(calculated using $\alpha_i^{r}$ and $\beta_i^{r}$
which were found previously
to fit the attenuation along a single ray).
$\Gamma_i^{sp}$ is a simple, though approximate, expression
for the photodissociation rate at a point $Q$ interior to a sphere,
and $\alpha_i^{sp}$ and $\beta_i^{sp}$ are the best
fit attenuation parameters for points interior to a sphere.

In the particular situation of a circumstellar envelope, 
we note that most of the
photodissociation of H$_2$O and OH occurs at low $A_v$, near
the surface of the sphere, and thus it is most important that
$\Gamma_i^{sp}$ be accurate for relatively small $A_v$.  To that end,
we fit our own bi-exponential to the tabulated photodissociation 
rates at small $A_v$ from R91,
which yielded the $\alpha_i^{pp}$ and $\beta_i^{pp}$, listed
in Table \ref{photorates}.  We then applied
the fitting procedure described above and found values for 
$\alpha_i^{r}$, $\beta_i^{r}$, $\alpha_i^{sp}$ and $\beta_i^{sp}$,
also listed in Table \ref{photorates}.

\clearpage

\begin{deluxetable}{c c c c c }
\tablecaption{Line Fit Parameters And Upper Limits \label{linefit}}
\tablehead{\colhead{Species} 
& \colhead{Frequency} 
& \colhead{Line Strength}
& \colhead{Line Center} 
& \colhead{Line Width} \\
& \colhead{(MHz)}
& \colhead{(mK km s$^{-1}$)}
& \colhead{($v_{LSR}$ in km s$^{-1}$)}
& \colhead{(FWHM in km s$^{-1}$)}
}
\tablewidth{0pt}

\startdata
H$_2$O\tablenotemark{a} & 556936 & 414 $\pm$ 17 & $-24.7$ $\pm$ 0.2 & 24.2 $\pm$ 0.2\\
OH & 1667 & 39.0 $\pm$ 4.45 & $-36.1$ $\pm$ 0.34 & 5.8 $\pm$ 0.71\\
OH & 1665 & 23.7 $\pm$ 5.05 & $-38.7$ $\pm$ 0.58 & 5.8 $\pm$ 1.52\\
OH & 1612 & $<$22.6 ($3\sigma$)\tablenotemark{b} & & \\
HCN & 1347 & $<$48.9 ($3\sigma$)\tablenotemark{c} & & \\
\enddata
\tablecomments{We list the line fit parameters for spectra displayed
in Figure \ref{spec}, as well as the upper limits on other observed
frequencies where no lines were detected.}
\tablenotetext{a}{These values correct a small error in the fit to the
SWAS spectrum reported by Melnick et al. (2001).  The dashed line in
Figure \ref{spec} is a plot of the corrected fit.  
The fitted values assume a parabolic line shape, and
the quoted errors are formal errors derived from a 3 parameter fit.
Due to the uncertainty in the intrinsic 
line shape, a more appropriate estimate
of the velocity errors is simply the size 
of a velocity resolution element,
$1.1\,{\rm km\,s^{-1}}$.  The $v_{LSR}$ of IRC+10216
measured by the water vapor data is
therefore consistent with previous measurements of the
$v_{LSR}$ of the source.}
\tablenotetext{b}{Assumes a line width of $5.8\,$km s$^{-1}$.}
\tablenotetext{c}{Assumes a line width of $30\,$km s$^{-1}$.}
\end{deluxetable}

\clearpage

\begin{deluxetable}{c c c c}
\tablecaption{Photodissociation Models \label{phottab}}
\tablehead{\colhead{Model} 
& \colhead{$\dot M$} 
& \colhead{$G_0$}
& \colhead{${A_v/N_H \over {(A_v/N_H)_0}}$} \\
& \colhead{$10^{-5} {\rm (M_{\odot}\, yr^{-1})}$}
&
&
}
\tablewidth{0pt}

\startdata
A&3&1&1 \\
B&5&1&1 \\
C&3&1/3&1 \\
D&3&1&2 \\
E&3&1/3&9 \\
\enddata
\end{deluxetable}

\clearpage

\begin{deluxetable}{c c c c c}
\tablecaption{Line Profile Models \label{linetab}}
\tablehead{\colhead{Model} 
& \colhead{Photodissociation}
& \colhead{Opening angle, $\psi$}
& \colhead{b}
& \colhead{Flux}
\\
& \colhead{Model}
& \colhead{(degrees)}
& \colhead{${\rm (km\, s^{-1})}$}
& \colhead{${\rm (mJy\, km\, s^{-1})}$}
}
\tablewidth{0pt}

\startdata
S1&A&180&0.65& 10.5\\
S2&C&180&0.65& 18.1\\
S3&E&180&0.65& 23.9\\
R1&A&10 &0.65& 10.6\\
R2&A&1  &0.65& 10.6\\
R3&A&0.1&2.0 & 10.6\\
\enddata
\end{deluxetable}

\clearpage

\begin{deluxetable}{c c c c c c c c}
\tablecaption{$\alpha_i$ and $\beta_i$ Coefficients
for H$_2$O and OH. \label{photorates}}
\tablehead{\colhead{Species}
& \colhead{$C\; {\rm (s^{-1})}$}
& \colhead{$\alpha^{pp}$} 
& \colhead{$\beta^{pp}$}
& \colhead{$\alpha^{r}$}
& \colhead{$\beta^{r}$}
& \colhead{$\alpha^{sp}$}
& \colhead{$\beta^{sp}$}
}
\tablewidth{0pt}

\startdata
H$_2$O&$3.19\times 10^{-10}$&3.03&$-0.180$&1.73&0.00&2.22&0.00\\
OH    &$1.90\times 10^{-10}$&3.13&$-0.195$&1.77&0.00&2.26&0.00\\
\enddata
\end{deluxetable}

\clearpage

\begin{references}
\reference{and92} Andersson, B.-G., Wannier, P.G., \& Lewis, B.M.
1992, MNRAS, 254, 7
\reference{cer00} Cernicharo, J., Gu\'elin, M., \& Kahane, C. 2000,
A\&AS, 142, 181
\reference{cha91} Chapman, J. M., Cohen, R. J., \& Saikia, D. J. 
1991, MNRAS, 229, 247
\reference{dot97} Doty, S. D., \& Leung, C. M. 1997, MNRAS, 286, 1003
\reference{dra78} Draine, B. T. 1978, ApJS, 36, 595
\reference{eli78} Elitzur, M. 1978, A\&A, 62, 305
\reference{eng94} Engels, D. 1994, A\&A, 285, 497
\reference{eto01} Etoka, S., Blaszkiewicz, L., Szymczak, M., \&
Le Squeren, A.M. 2001, A\&A, 378, 522
\reference{fon03} Fong, D., Meixner, M., \& Shah, R. Y. 2003, ApJ, 582, L39
\reference{for01} Ford, K.E.S., \& Neufeld, D.A. 2001, ApJ, 557, L113
\reference{gro98} Groenewegen, M. A. T., Van Der Veen, W. E. C. J., 
\& Matthews, H. E. 1998, A\&A, 338, 491
\reference{gla96} Glassgold, A. E. 1996, ARA\&A, 34, 241
\reference{hei99} Heiles, C. 1999, Arecibo Technical 
and Operations Memo 1999-02
\reference{jur83} Jura, M. 1983, ApJ, 267, 647
\reference{leb97} LeBertre, T. 1997, A\&A, 324, 1059
\reference{luc89} Lucas, R., \& Cernicharo, J. 1989, A\&A, 218, L20
\reference{luc92} Lucas, R., \& Guilloteau, S. 1992, A\&A, 259, L23
\reference{mau99} Mauron, N., \& Huggins, P. J. 1999, A\&A, 349, 203
\reference{mau00} Mauron, N., \& Huggins, P. J. 2000, A\&A, 359, 707
\reference{mel01} Melnick, G. J., Neufeld, D. A., 
Ford, K. E. S., Hollenbach, D. J., \& Ashby, M. L. N. 2001, Nature, 412, 160
\reference{men02} Men'shchikov, A. B., Hofmann, K.-H., \&
Weigelt, G. 2002, A\&A, 392, 921
\reference{rob91} Roberge, W.G., Jones, D., Lepp, S., \& Dalgarno, A.
1991, ApJS, 77, 287
\reference{sah89} Sahai, R., Claussen, M. J., \& Masson, C. R. 
1989, A\&A, 220, 92
\reference{ski95} Skinner, C. J., Justtanont, K., Tielens, A. G. G. M., 
Betz, A. L., \& Boreiko, R. T. 1995, ASP Conf. Ser. 73, Airborne 
Astronomy Symp. on the Galactic Ecosystem:
From Gas to Stars to Dust, ed. M. R. Haas, J. A. Davidson, 
\& E. F. Erickson (San Francisco: ASP), 433
\reference{ski99} Skinner, C. J., Justtanont, K., Tielens, A. G. G. M., 
Betz, A. L., Boreiko, R. T., \& Baas, F. 1999, MNRAS, 302, 293
\reference{tut00} Tuthill, P. G., Monnier, J. D., Danchi, W. C.,
\& Lopez, B. 2000, ApJ, 543, 284
\reference{wal98} Wallerstein, G., \& Knapp, G. R. 1998, ARA\&A, 36, 369
\reference{wei02} Weigelt, G., Balega, Y. Y., Bl\"ocker, T.,
Hofmann, K.-H., Men'shchikov, A. B., \& Winters, J. M. 2002, A\&A, 392, 131
\reference{win94} Winters, J. M., Dominik, C., 
\& Sedlmayr, E. 1994, A\&A, 288, 255
\end{references}
\end{document}